\documentclass{article}
\usepackage[final,nonatbib]{neurips_2020}

\usepackage{neurips_2020}

\usepackage[bibstyle=ieee,citestyle=numeric-comp,sorting=nyt]{biblatex}

\usepackage{biblatex}

\usepackage[utf8]{inputenc} 
\usepackage[T1]{fontenc}    
\usepackage{hyperref}       
\usepackage{url}            
\usepackage{booktabs}       
\usepackage{amsfonts}       
\usepackage{nicefrac}       
\usepackage{xcolor}
\usepackage{microtype}      
\DeclareUnicodeCharacter{0301}{}
\title{Anticipatory Ethics and the Role of Uncertainty}

\author{%
 Priyanka Nanayakkara \quad Nicholas Diakopoulos \quad Jessica Hullman \\
  Northwestern University\\
  \texttt{priyankan@u.northwestern.edu, \{nad,jhullman\}@northwestern.edu}
}

\begin{document}

\maketitle

\begin{abstract}
Making conjectures about future consequences of a technology is an exercise in trying to reduce various forms of uncertainty. Both to produce and reason about these conjectures requires understanding their potential limitations. In other words, we need systematic ways of considering uncertainty associated with given conjectures for downstream consequences. In this work, we frame the task of considering future consequences as an anticipatory ethics problem, where the goal is to develop scenarios that reflect plausible outcomes and their ethical implications following a technology’s introduction into society. In order to shed light on how various forms of uncertainty might inform how we reason about a resulting scenario, we provide a characterization of the types of uncertainty that arise in a potential scenario-building process.
\end{abstract}

\section{Introduction}
Researchers have called attention to the ways in which technology can have undesirable consequences for society, in particular by reinforcing or exacerbating inequalities that impact marginalized people~\cite{eubanks2018automating,benjamin2019race,noble2018algorithms}. Calls from within the computer science research community propose making data and models more transparent~\cite{gebru2018datasheets,mitchell2019model}, argue for more discussion around ethical implications of research within the community itself~\cite{bruckman2020have}, and suggest that discussion of potential negative consequences ought to be incorporated into the peer-review process~\cite{hecht2018s} (similar to the NeurIPS ``broader impact'' statement).

In considering broader impacts, researchers must take into account many factors---each of which is characterized by uncertainty---that could influence how a given technology impacts society. For example, researchers must conceptualize broad societal factors such as social, economic, legal, and political conditions, as well as who might be likely to use a technology and that actor/organization’s motivations. Faced with a task as daunting as predicting future outcomes, researchers may be quick to write the task off as entirely impossible, seeing it as an exercise in making arbitrary guesses about the future. However, distinguishing between types of uncertainty that arise in the process of considering future outcomes can help the computer science community identify ways that researchers can approach thinking through future outcomes by reducing certain forms of uncertainty.

Drawing on analogies to statistical inference, we distinguish between multiple forms of uncertainty that may enter the process of considering future impacts. We intend for this work to 1) help people reason about stated future consequences, 2) motivate the need for researchers to provide insight into how they thought through these consequences, and 3) expose types of uncertainty that might be reduced.

We begin by describing anticipatory governance and ethics---an approach to articulating the social and ethical implications of emerging technologies such that future developments may be steered~\cite{guston2014understanding}. Then, we describe different types of statistical uncertainty. Finally, we characterize uncertainty that arises specifically in the context of scenario building, a common method used to produce foresight for anticipatory ethics approaches. 

\section{Background}
We first provide a brief overview of anticipatory ethics in order to contextualize our specific focus on explorative scenario building. We then outline a typology of uncertainty motivated by the practice of statistical modeling. Finally, we apply this typology to the task of explorative scenario building.

\textbf{Anticipatory Ethics.} Anticipatory ethics is a form of anticipatory governance focused on helping designers and others to articulate the ethical implications of technology. It can incorporate various methods that integrate ethical analysis with foresight analysis to assess future ethical impacts of emerging technologies~\cite{brey2017ethics}. For instance, Delphi panels ask experts to fill out questionnaires around anticipated future outcomes, then ask them to revise their responses based on summaries of the group’s answers, and finally aggregate these opinions or scores~\cite{brey2017ethics}. This method primarily relies on expert opinions to converge to an agreed-upon anticipated future that will result from a particular technology. Other anticipatory ethics methods, such as participatory foresight~\cite{nikolova2014rise}, include ways of involving citizens in the process to provide feedback around which type of future outcome they would like to realize~\cite{brey2017ethics}.

We are particularly interested in focusing on scenario methods within anticipatory ethics, as they seem most relevant to researchers writing broader impact statements. In general, scenario methods are ways of producing narrative-style descriptions of future possibilities. Broadly, these methods fall into three categories: predictive (“what will happen?”), explorative (“what could happen?”), and normative (“how can a certain target be reached?”)~\cite{borjeson2006scenario}. Here, we focus on explorative scenario building. For our purposes, we are not interested in how researchers might begin to place mathematical probabilities on various outcomes (i.e., predictive scenarios). Rather, we are interested in how they might qualitatively consider a wide range of ``plausible” scenarios~\cite{ramirez2014plausibility}, as these scenarios could serve as the basis for well-thought-out broader impact statements.

As one example application of the anticipatory ethics approach using exploratory scenarios, Diakopoulos and Johnson~\cite{diakopoulos2019anticipating} studied ethical issues surrounding deepfakes (synthetic audio/video of real people) in an election context. They created a typology of conceptual dimensions (e.g., actors, motivations, channels for distribution, phases of a campaign) that might be included in a scenario, and are applicable to an election context. In each scenario, conceptual dimensions took on specific values (e.g., the dimension ``actors’’ could take on values of ``candidate,’’ ``campaign staff,’’ etc.). Thus, they scoped resulting scenarios by grounding them both in real-world contexts and in the current capabilities of the technology. Importantly, the quality of resulting scenarios hinges on several factors, including how comprehensive the conceptual dimensions are in allowing for thorough investigation of a technology’s impact within sociotechnical systems, as well as uncertainty surrounding the possibility, plausibility, feasibility, or probability of particular scenarios being realized.

\textbf{Typology of Uncertainty.} While there is not one single agreed upon typology for describing types of uncertainty within a statistical modeling framework, we will focus on the relatively prevalent typology of aleatory, epistemic, and ontological uncertainty~\cite{spiegelhalter2017risk}. 

While epistemic uncertainty describes limitations in confidence about existing knowledge~\cite{fox2011distinguishing}, aleatory uncertainty describes uncertainty ``due to the fundamental indeterminacy or randomness in the world’’~\cite{van2019communicating}. As Fox and {\"U}lk{\"u}men~\cite{fox2011distinguishing} put it, ``[a]leatory uncertainty is attributed to outcomes that for practical purposes cannot be predicted and are therefore treated as stochastic (e.g., the result of a coin flip), whereas epistemic uncertainty is attributed to missing information or expertise (e.g., whether or not one has correctly answered a question on an exam)...’’ While we cannot reduce aleatory uncertainty, we can reduce epistemic uncertainty by gathering more information on ``past or present phenomena that we currently don’t know but could, at least in theory, know or establish’’~\cite{van2019communicating}.

Ontological uncertainty is the most philosophical in nature, describing ``uncertainty about the entire modeling process as a description of reality’’~\cite{spiegelhalter2017risk}. To better understand ontological uncertainty, it is useful to more broadly consider the concept of ``large world’’ uncertainty (versus ``small world’’ uncertainty)~\cite{mcelreath2020statistical}. We can think of the ``small world’’ as a statistical model, a necessarily limited representation of the real world (i.e., the ``large world’’). Thus, small world uncertainty describes uncertainty around the range of output possibilities, and their corresponding probabilities, assuming all the model’s assumptions are correct. Large world---or ontological---uncertainty acknowledges the extent to which the model may not even contain adequate information to properly model the large world.

\section{Characterizing Uncertainty in Scenario Building}
We envision a situation in which a team of computer science researchers have just created a new technology, and are trying to anticipate future consequences of their work. In this situation, the researchers want to produce a written discussion of both potential positive and negative future consequences, perhaps for the purposes of a NeurIPS-style broader impact statement. 

Thus, we imagine the researchers creating ``scenarios’’ to explore a range of future impacts (in an anticipatory ethics sense) and ultimately using the ethical issues that become salient through these scenarios as the basis for writing their discussion on consequences. As previously mentioned, we are focused on researchers engaging in explorative scenario building as a tool to qualitatively consider a wide range of plausible outcomes once a technology has already been built and its capabilities have become somewhat clear. 

We make the analogy between scenario building and statistical modeling to characterize uncertainty that arises in the scenario building process, and uncertainty that is baked into resulting scenarios. The team of researchers can be thought of as building a ``model of society’’ (i.e., a ``small world'') by first choosing to include various ``parameters’’ in the model. These might include what Diakopoulos and Johnson~\cite{diakopoulos2019anticipating} refer to as ``conceptual dimensions’’ (e.g., actors, motivations, channels for distribution, phases of a campaign, etc.) plus factors meant to surface broader concepts such as social, legal, economic, or political conditions which could inform how a technology is likely to be used~\cite{uruena2019understanding} (e.g., economic systems such as capitalism, political systems such as democracy or authoritarianism, histories of discrimination on the basis of race, etc.). Note that these broader concepts are highly complex, and might require several interrelated ``parameters'' in the model to adequately surface relevant behaviors related to them. Next, researchers ``estimate’’ the ``values’’ of these parameters by relying on their existing knowledge around how these parameters might behave in the real world or by seeking out new knowledge to fill gaps in their understanding. It is important to note that this process is distinct from, and more subjective than, parameter estimation in a strictly statistical sense (e.g., least-squares method for estimating parameters in a linear regression model), since it is not necessarily governed by any underlying computational philosophy. However, the more data-driven the parameter estimation process can be, the closer the analogy to statistical inference. Finally, researchers use the ``model,’’ complete with their parameter ``estimates,’’ to generate future scenarios/consequences using anticipatory ethics methods. This subjective inference process yields a set of scenarios in which some scenarios are deemed more likely than others, based on the researchers’ prior knowledge about the world, which is encoded into their ``estimates’’ of parameters.

For the following sections, we rely on Diakopoulos and Johnson’s~\cite{diakopoulos2019anticipating} work anticipating ethical issues around deepfakes in an election context to ground examples of uncertainty that generally arise in the process of researchers creating scenarios.

\textbf{Epistemic Uncertainty.} There is epistemic uncertainty around parameter values. For instance, deepfakes could potentially spread through social media. The model’s parameter for social media is only as good as, for instance, how well it captures the ways in which false information spreads more rapidly than true information online~\cite{vosoughi2018spread}, or, for instance, the ways in which women may be especially targeted for online harassment or intimidation through deepfake pornography. If the researchers estimate the social media parameter by encoding the idea that \textit{true} information spreads more quickly online, then resulting scenarios from the model will be correspondingly off. Thus, the researchers’ poor estimates of parameters will impact resulting output scenarios. If the researchers are unable to adequately consider real-world factors, then the scenarios will not represent what we might see in the real world either. Furthermore, there is uncertainty around how well the researchers are able to capture interactions between parameters (e.g., how candidates use social media). 

Thus, ``estimating’’ parameters is a crucial task. How well parameters are estimated hinges on how informed researchers are in a particular area or context of use and whether they have a grasp over existing issues of concern (e.g., disparities that are likely to be worsened). To better estimate a parameter’s value, researchers might lean on domain experts and research from relevant disciplines. In the case of the latter, and in the example of a ``candidate-actor’’ parameter, it might be useful to consult with people who have run for public office to better understand the specific challenges they faced while campaigning. It could be, for instance, that online trolling was a challenge~\cite{nytimesfemalecandidates}, and this could inform scenarios around how deepfakes could be used as part of a variation on previous forms of harassment. Additionally, participatory approaches that involve collecting feedback from a diverse range of people could help to better estimate parameters, and to include a more comprehensive understanding of how a certain parameter in society is likely to operate. 

Decisions researchers make to estimate parameters---whether and how they consult with domain experts, engage in participatory methods, etc.---can be likened to ``researcher degrees of freedom’’~\cite{simmons2011false}. These decisions introduce hard to quantify uncertainty, as there are numerous ways in which one might go about ``estimating'' these parameters. Additionally, given incentives to produce discoveries that are deemed to be clearly useful, researchers may be more inclined to anticipate more positive outcomes of their work for society than negative; this predisposition may come through in the ways they estimate parameters. This raises the question of whether researchers are well-positioned to do this type of post-research, explorative scenario building, or if this work might more effectively be done by (or in collaboration with) specialists who understand the technology and its modes of action and who have a well-considered and researched model of society.

\textbf{Ontological Uncertainty.} Ontological uncertainty arises in the inclusion and exclusion of various parameters. In other words, whether researchers pick the ``right’’ set of parameters to include in a model is ontologically uncertain. For instance, if the researchers do not consider social media in the model to consider election deepfakes, they cannot even consider scenarios where deepfakes play a role in online social spaces. Researchers may reduce ontological uncertainty by relying on perspectives and ideas from other fields; these perspectives may allow them to see the world through different lenses, and therefore open up a wider range of possibilities. For example, considering the ways in which histories of colonialism shape current geopolitical and cultural power dynamics (e.g., in the context of artificial intelligence~\cite{mohamed2020decolonial}) could help widen the set of relevant parameters.

Furthermore, ontological uncertainty arises around philosophical questions of what constitutes a positive or negative consequence. While outcomes that are, for instance, discriminatory on the basis of protected attributes, are clearly negative consequences, there may not be agreement around one particular definition of fairness~\cite{corbett2018measure,hutchinson201950,saxena2019fairness} or which definitions apply in various contexts~\cite{verma2018fairness}. Therefore, it may be ontologically uncertain which definitions should be prioritized in the model.

\textbf{Aleatory Uncertainty.} There is aleatory uncertainty around resulting scenarios due to irreducible noise stemming from the randomness of natural processes and the unpredictability and variability of human behavior. Researchers cannot reduce this uncertainty ahead of time, no matter how much extra information they gather. For instance, chance may play a role in the success of a cultural product like a book or movie~\cite{salganik2009web}. In the case of a deepfake, chance may play a role in whether a deepfake goes viral. Resulting scenarios should therefore be read with aleatory uncertainty in mind; that is, there is some random noise in whether the scenarios will be realized that can’t necessarily be mapped back to concrete factors, such as the choice of parameters or their estimates.

\section{Conclusion}
\vspace{-.5mm}
By categorizing multiple forms of uncertainty and showing how they apply in explorative scenario building, we show that recognizing distinctions between different types of uncertainty inherent in forecasting may help researchers produce more meaningful broader impact statements. In addition, this work motivates the need for researchers to disclose how they scoped the task of considering future impacts and how they generally thought through future consequences (e.g., how they chose ``parameters’’). These explanations will allow us insight into the uncertainty implicit in a set of scenarios, and therefore insight into the strengths and limitations of both the scenarios and implied broader impacts. This work also suggests that both a breadth and depth of expertise and knowledge may be important to reducing both epistemic and ontological uncertainty, something that could be supported through participatory methods, or collaborations with other specialists or ethicists. An interesting question for future work is how researchers search the space of possible models of societal dynamics proposed and how ``sticky'' models are in the sense of anchoring subsequent research (for analogies to statistical inference in science, see~\cite{devezer2019scientific}). Finally, this work suggests that more work remains to be done to develop and teach systematic methodologies (e.g., anticipatory ethics, including scenario building with careful consideration of uncertainty) for broader impact writing.

\printbibliography
\end{document}